\documentclass[aps,reprint]{revtex4-1}

\usepackage{graphicx}
\usepackage{array}
\usepackage{bm}
\usepackage[english]{babel}
\usepackage{dcolumn}
\usepackage{amssymb}
\usepackage{amsmath}

\begin{document}

\title{Approximate solutions for $\bm{N}$-body Hamiltonians with identical particles in $\bm{D}$ dimensions}

\author{Claude \surname{Semay}}
\email[E-mail: ]{claude.semay@umons.ac.be}
\author{Christophe \surname{Roland}}
\email[E-mail: ]{christophe.roland@alumni.umons.ac.be}
\affiliation{Service de Physique Nucl\'{e}aire et Subnucl\'{e}aire,
Universit\'{e} de Mons, UMONS Research Institute for Complex Systems,
Place du Parc 20, 7000 Mons, Belgium}

\date{\today}

\begin{abstract}
A method based on the envelope theory is presented to compute approximate solutions for $N$-body Hamiltonians with identical particles in $D$ dimensions ($D\ge 2$). In some favorable cases, the approximate eigenvalues can be analytically determined and can be lower or upper bounds. The accuracy of the method is tested with several examples, and an application to a $N$-body system with a minimal length is studied. Finally, a semiclassical interpretation is given for the generic formula of the eigenvalues. 
\end{abstract}

\pacs{03.65.Ge,03.65.Sq}
%03.65.Ge Solutions of wave equations: bound states
%03.65.Sq Semiclassical theories and applications

\maketitle

\section{Introduction}
\label{sec:intro} 

The quantum-mechanical $N$-body problem is highly nontrivial as soon as $N>2$. Because it arises in all areas where quantum mechanics is involved, from atomic to hadronic physics, the amount of papers devoted to that topic is huge. Some useful results can be found in several textbooks \cite{flug,fett,suth,kinetic,olsh}. Among all the possible techniques, the envelope theory, also known as the auxiliary field method, is a powerful method to obtain approximate solutions, eigenvalues and eigenstates, of eigenequations in quantum mechanics \cite{buis09}. The basic idea is to replace the Hamiltonian $H$ under study by an auxiliary Hamiltonian $\widetilde H$ which is solvable, the eigenvalues of $\widetilde H$ being optimized to be as close as possible to those of $H$. This method has been used to tackle relativistic and semirelativistic systems of identical particle in the three dimensional space \cite{hall88,hall06,silv10}. Recently the envelope theory has been extended to treat one-body and two-body problems with an arbitrary kinematics \cite{sema12}. 

The purpose of this paper is to generalize the envelope theory to solve approximatively the $N$-body problem for identical particles with arbitrary kinetics in a $D$ dimensional space ($D\ge 2$). This work is motivated by the existence of non-standard kinetic energies in some physical problems, for instance in atomic physics with non-parabolic dispersion relation \cite{arie92}, in hadronic physics  with particle masses depending on the relative momentum \cite{szcz96}, or in quantum mechanics with a minimal length \cite{kemp95,brau99,ques10,buis10b}. Moreover, problems in $D$ dimensions can appear in various physical situations. In particular, $D=2$ systems can be used as toy models for $D=3$ systems \cite{tepe99} or are the natural framework for the physics of anyons \cite{murt91,khar05}. So, the possible domains of interest for the method are numerous.

This paper is organized as follows. The one-body and two-body cases deserve a special treatment. They are presented together in Section~\ref{sec:12}. The $N$-body case is studied in Section~\ref{sec:n} where some properties of the solutions are presented. The accuracy of the method is tested with several examples in Section~\ref{sec:app}, where an application to a $N$-body system with a minimal length is studied. Concluding remarks are given in Section~\ref{sec:rem}. A semiclassical interpretation of the generic formula obtained for the eigenvalues is presented in the Appendix. Let us mention that the results obtained here are quite direct generalizations of those given in Refs.~\cite{silv10,sema12}, where most of the details about the calculations can be found. These results, which can be useful to a large community, are not presented elsewhere.

\section{One- and two-body cases}
\label{sec:12} 

Let us assume that the Hamiltonian $H$ can be written as (in the following, we will work in natural units $\hbar = c = 1$)
\begin{equation}
\label{TpV}
H = T(p) + V(r),
\end{equation}
with $p=|\bm p|$ and $r=|\bm r|$, $\bm r$ and $\bm p$ being conjugate variables. For the one-body case, $\bm r$ is the distance of the particle from the center of force. For the two-body case, $\bm r$ is the distance between the particles. In Ref.~\cite{sema12}, the procedure to compute the approximate solutions is based on a good parameterization for $D=3$ of the eigenvalues $E_0$ of the Hamiltonian $H_0$ (which differs from the auxiliary Hamiltonian $\widetilde H$ by a constant term) with the auxiliary potential $P(r)$,
\begin{equation}
\label{H0}
H_0 =  \frac{\bm p^2}{2 \mu}+ \rho\,P(r)  \quad \textrm{with} \quad P(r)=\textrm{sgn}(\lambda)\, r^\lambda
\end{equation}
($\mu>0$, $\rho >0$ and $0 \ne \lambda > -2$). We have 
\begin{equation}
\label{epsPr}
E_0 = \frac{\lambda+2}{2 \lambda}\left( |\lambda|\rho \right)^{2/(\lambda+2)}\left(\frac{Q^2}{\mu}\right)^{\lambda/(\lambda+2)}, 
\end{equation}
where $Q$ is a global quantum number. This expression is also relevant for arbitrary values of $D\ge 2$. The value of $Q$ is exactly known only for the Coulomb interaction ($\lambda=-1$, $Q=n+l+\frac{D-1}{2}$) and the harmonic potential ($\lambda=2$, $Q=2\, n+l+\frac{D}{2}$) \cite{yane94,aver88}. If $\lambda=1$, $Q$ is known only in the case of $D=3$ for $l=0$ states and is equal to $2 (-\alpha_n/3)^{3/2}$, where $\alpha_n$ is the $(n + 1)$th zero of the Airy function Ai. In all other cases, (\ref{epsPr}) can considered as a definition for $Q$. 

Following the same procedure as the one described in Ref.~\cite{sema12}, the approximate eigenvalue $E$ is given by the following set of equations:
\begin{eqnarray}
\label{AFM1}
&&E = T(p_0)+V(r_0), \\
\label{AFM2}
&&r_0\,p_0 = Q, \\
\label{AFM3}
&&p_0\, T'(p_0) = r_0\, V'(r_0).
\end{eqnarray}
The parameter $r_0$ can be interpreted as the mean distance between the particles and $p_0$ as the mean momentum per particle \cite{sema12}. Let us note that (\ref{AFM3}) is the translation into the variables $r_0$ and $p_0$ of the generalized virial theorem \cite{virial}. From the symmetry of equations~(\ref{AFM1})-(\ref{AFM3}) under the swap of $p_0$ and $r_0$ variables, it is clear that a Hamiltonian and its Fourier transform are characterized by the same solutions.

In some cases, the approximate solution $E$ can be a lower or an upper bound. Let us define two functions $b_T$ and $b_V$ such that 
\begin{equation}
\label{hg}
T(x) = b_T(x^2) \quad \textrm{and}\quad V(x) = b_V(P(x)).
\end{equation}
If $b_T''(x)$ and $b_V''(x)$ are both concave (convex) functions, $E$ is an upper (lower) bound of the genuine eigenvalue \cite{sema12}. If $T(p) \propto p^2$ ($V(r) \propto P(r)$), the variational character is solely ruled by the convexity of $b_V(x)$ ($b_T(x)$). In the other cases, the variational character of the solution cannot be guaranteed. Since many techniques exist to compute accurate numerical solutions of one- or two-body problems, this method is only interesting if the system~(\ref{AFM1})-(\ref{AFM3}) allows an analytical solution.

\section{$\bm{N}$-body case}
\label{sec:n} 

Let us now consider the $N$-body Hamiltonian for identical particles, in a $D$ dimensional space, interacting via the one-body $U$ and two-body $V$ interactions
\begin{equation}
\label{HNb}
H=\sum_{i=1}^N T(|\bm p_i|) + \sum_{i=1}^N U\left(|\bm r_i - \bm R|\right) + \sum_{i\le j=1}^N V\left(|\bm r_i - \bm r_j|\right),
\end{equation}
where $\sum_{i=1}^N \bm p_i = \bm 0$ and $\bm R = \frac{1}{N}\sum_{i=1}^N \bm r_i$ is the center of mass position. A one-body potential such as $U$ is sometimes used to simulate confinement in hadronic systems \cite{buis10}. The possibility to compute an approximate solution with the envelope theory relies on the existence of a completely soluble $N$-body auxiliary Hamiltonian. This is the case for the $N$ identical oscillator Hamiltonian
\begin{equation}
\label{H0N}
H_0=\frac{1}{2\mu}\sum_{i=1}^N \bm p_i^2 + \nu \sum_{i=1}^N \left(\bm r_i - \bm R\right)^2 + \rho \sum_{i\le j=1}^N \left(\bm r_i - \bm r_j\right)^2.
\end{equation}
The complete solution is given in Ref.~\cite{silv10} for $D=3$, but the result is easily generalizable to any values of $D$. Following the same procedure as the one given in Ref.~\cite{silv10}, but introducing an auxiliary counterpart for the kinetic part as in Ref.~\cite{sema12}, the approximate eigenvalue $E$ is given by the following set of equations for a completely (anti)symmetrized state:
\begin{eqnarray}
\label{AFM1N}
&&E=N\, T(p_0) + N\, U \left( \frac{r_0}{N} \right) + C_N\, V \left( \frac{r_0}{\sqrt{C_N}} \right), \\
\label{AFM2N}
&&r_0\, p_0=Q, \\
\label{AFM3N}
&&N\, p_0\, T'(p_0) =  r_0\, U' \left( \frac{r_0}{N} \right) + \sqrt{C_N}\, r_0\, V' \left( \frac{r_0}{\sqrt{C_N}} \right),
\end{eqnarray}
where $C_N=N(N-1)/2$ is the number of particle pairs and where
\begin{equation}
\label{QN}
Q = \sum_{i=1}^{N-1} (2 n_i + l_i) + (N - 1)\frac{D}{2}.
\end{equation}
The approximate eigenstate is given in terms of harmonic oscillator functions \cite{silv12}.

As $E$ depends on $Q$, this method does not raise the strong degeneracy inherent to $H_0$ (see also (\ref{AFM2})). Following the nature of the particles, only some set of quantum numbers are allowed. For instance, the ground state of $N$ bosons is given by $Q_{\textrm{GS}}^{\textrm{B}} = (N - 1)\frac{D}{2}$. For fermions, the calculation is more involved. For a large number of particles, on can find
\begin{equation}
\label{QNferm}
\lim_{N\to\infty} Q_{\textrm{GS}}^{\textrm{F}}= \frac{D}{D+1} \left( \frac{D!\, N^{D+1}}{d} \right)^{1/D}, 
\end{equation}
where $d$ is the degeneracy of the fermion. 

Equations~(\ref{AFM1N})-(\ref{QNferm}) were obtained in Ref.~\cite{silv10,silv11}, but only for $D=3$ and for $T(p) \propto p^2$ or $\sqrt{p^2+m^2}$. Other calculations performed in these last references are immediately generalizable to arbitrary $D$ and $T(p)$. We give here the results for the existence of bounds, the presence of perturbative interactions and the definition of critical coupling constants.

Let us now define three functions $b_T$, $b_U$ and $b_V$ such that 
\begin{equation}
\label{hg}
T(x) = b_T(x^2), \  U(x) = b_U(x^2) \  \textrm{and} \  V(x) = b_V(x^2).
\end{equation}
If $b_T''(x)$, $b_U''(x)$ and $b_V''(x)$ are all concave (convex) functions, $E$ is an upper (lower) bound of the genuine eigenvalue. If the second derivative is vanishing for one or two of these functions, the variational character is solely ruled by the convexity of the other(s). In the other cases, the variational character of the solution cannot be guaranteed. If no analytical solution can be found for the system~(\ref{AFM1N})-(\ref{AFM3N}), a numerical solution is easy to compute. Such approximation is interesting to obtain since an accurate numerical solution is always hard to compute for $N>2$ \cite{silv85,silv07}.  

Let us assume that the approximate solution $E$ is obtained for the $T$, $U$ and $V$ energy terms with the value $r_0$, and that these terms are respectively supplemented by perturbations $\tau\, t(p) \ll T(p)$, $\eta\, u(x) \ll U(x)$ and $\epsilon \, v(x) \ll V(x)$ ($\tau$, $\eta$ and $\epsilon$ are small parameters) in the physical domain of interest. Then, one can show that the approximate solution $E_p$, at first order in $\tau$, $\eta$ and $\epsilon$ is given by \cite{silv11}
\begin{equation}
\label{Epert}
E_p=E 
+ N\, \tau\, t \left( p_0 \right) 
+ N\, \eta\, u \left( \frac{r_0}{N} \right) 
+ C_N\, \epsilon\, v \left( \frac{r_0}{\sqrt{C_N}} \right).
\end{equation}
This result could seem quite obvious, but it demonstrates that the knowledge of $r_0$ is sufficient to obtain the
contribution of the perturbations at the first order.

Some interactions, as the Yukawa or the exponential potentials, admit only a finite number of bound states. They can be written under the form $W(x)=-\kappa\, w(x)$, where $\kappa$ is a positive quantity which has the dimension of an energy and $w(x)$ a ``globally positive" dimensionless function vanishing at infinity. The critical coupling constant $\kappa_c(\{\theta\})$, where $\{\theta\}$ stands for a set of quantum numbers, is such that the potential admits a bound state with the quantum numbers $\{\theta\}$ if $\kappa > \kappa_c(\{\theta\})$ (see for instance Refs.~\cite{brau2,brau4}). 

Let us consider a nonrelativistic $N$-body system with particles of mass $m$, one-body potential $U(x)=-k\, u(x)$ or two-body potential $V(x)=-g\, v(x)$, both interactions admitting only a finite number of bound states. If the approximate eigenvalue of the energy is a lower (upper) bound, the approximate critical coupling constant is a lower (upper) bound of the genuine critical coupling constant. Assuming that only two-body forces are present, we obtain for the critical constant $g_c$
\begin{eqnarray}
\label{gc}
&&g_c  = \frac{1}{y_0^2\,v(y_0)} \frac{2}{N(N-1)^2} \frac{Q^2}{m}, \\
\label{y0gc}
&&2\, v(y_0)+y_0\, v'(y_0)=0.
\end{eqnarray}
With only one-body forces, a similar result is found for the critical constant $k_c$
\begin{eqnarray}
\label{kc}
&&k_c = \frac{1}{y_0^2\,u(y_0)} \frac{1}{2 N^2} \frac{Q^2}{m},
\\
\label{y0kc}
&&2\, u(y_0)+y_0\, u'(y_0)=0,
\end{eqnarray}
The variable $y_0$, determined by (\ref{y0gc}) or by (\ref{y0kc}), is independent of $N$, $Q$ (given by (\ref{QN})) and $m$, and depends only on the form of the function $v(x)$ or $u(x)$.

\section{Applications}
\label{sec:app} 

To have a lower or an upper bound is already a relevant information about an eigenvalue. But, one can ask if the bound is close or the not to the genuine value. This information is generally not obtained with the bound, and it is necessary to resort to comparisons with known solutions to test the accuracy of the bound. For the envelope theory developed here, it is convenient to examine separately the one/two-body cases from the many-body case. 

\subsection{One- and two-body cases}
\label{ssec:12} 

In the most favorable situations, both lower and upper bounds can be computed analytically for one/two-body Hamiltonians (see Section~\ref{sec:12}). For instance, this is the case for the square root potential, the logarithmic potential and some power-law potentials with a nonrelativistic kinematics. Lower and an upper bounds are computed for the dimensionless Hamiltonian $\bm p^2/4+\sqrt{r^2+\beta}$ for $D=3$ in Ref.~\cite{sema09}. For the lowest eigenvalues, the relative error on the upper bound is below 5\%. We have checked that a similar accuracy is obtained for $2 \le D \le 10$.

For other Hamiltonians, only one kind of bound can be computed. This is the case for the dimensionless Hamiltonian $\exp\left(k\, \bm p^2\right)+ r^2$, studied for $D=3$ in Ref.~\cite{sema12}. A very good lower bound can be computed for the three lowest eigenstates in the case $D=3$, for a range of the parameter $k$ varying from 0.001 to 1. We have checked that the accuracy is similar for $2 \le D \le 10$. 

In less favorable cases, no bound can be obtained, and only a direct comparison with numerical results can bring information about the accuracy of the method. Several Hamiltonians with nonrelativistic or semirelativistic kinematics are studied in Ref.~\cite{silv12} for $D=3$. Generally, a very good accuracy is obtained. Let us remark that, though the method is based on Hamiltonians $H_0$ with an infinite number of bound states, it works also well for Hamiltonians with a finite number of bound states.

\subsection{$\bm{N}$-body case}
\label{ssec:n} 

To obtain accurate numerical solutions in the $N$-body case is a challenging task. So, to test our results, we will also compare with other approximate methods. Let us recall that For $N >2$, only one kind of bound can be computed with the envelope theory (see Section~\ref{sec:n}).

The following Hamiltonian for three massless particles is relevant for the study of light baryons,
\begin{equation}
\label{hbaryon}
H=\sum^3_{i=1}\sqrt{\bm p^2_i}+a \sum^3_{i=1} |\bm r_i-\bm R|
-b \sum^3_{i<j=1}\frac{1}{|\bm r_i-\bm r_j|}.
\end{equation}
Accurate numerical eigenvalues for $D=3$ have been computed in Ref.~\cite{silv10} and compared with the upper bounds predicted by the envelope theory. The relative error on the lowest eigenvalues is around 10-20\%. Several improvements of the mass formula based on analytical procedures allow a reduction of this error to around 2\%, but the price to pay is the loss of the variational character of the formula.

Within the framework of the quantum chromodynamics with a great number of colors, the mass spectrum of the light baryons can be studied with a $N$-body generalization of (\ref{hbaryon}),
\begin{eqnarray}
\label{geneh}
\lefteqn{
H=\sum_{i=1}^{N}\left[\sqrt{\bm p^2_i}+a_1\left|\bm r_i-\bm R\right|\right] } \nonumber \\
&+&\sum_{i<j=1}^{N}\left[a_2\left|\bm r_i-\bm r_j\right|-\frac{b}{\left|\bm r_i-\bm r_j\right|}\right],
\end{eqnarray}
Using the envelope theory as in Ref.~\cite{buis10} but for $D$ dimensions, an upper bound $E_u$ of the ground state is given by
\begin{equation}
\label{genehup}
E_u^2 = 4\, C_N \left( a_1 + a_2 \sqrt{C_N} \right) \left( D - b \sqrt{C_N} \right).
\end{equation}
Using the method presented in Ref.~\cite{hall07}, a (quasi exact) lower bound $E_l$ can be computed \cite{buis10}
\begin{equation}
\label{genehlo}
E_l^2 = 2\, C_N \left( a_1 + a_2 N \right) \left( (D-1) - b (N-1) \right).
\end{equation}
It is difficult to estimate the accuracy of these bounds, but one can remark that $E_l$  and $E_u$ have the same behavior for large values of $N$ and $D$. So, we can have some confidence on the relevance of the upper bound computed with the envelope theory.

The Hamiltonian for a system of $N$ gravitating particles with a (nonrelativistic) Newton potential is given by
\begin{equation}
\label{hstar}
H=\sum^N_{i=1}\sqrt{\bm p^2_i+m^2} - \sum^N_{i<j=1}\frac{\alpha}{|\bm r_i-\bm r_j|},
\end{equation}
where $\alpha=G\, m^2$, $G$ being Newton's constant. It has been used to study boson stars \cite{rayn94}. Solving the system~(\ref{AFM1N})-(\ref{AFM3N}), we find an upper bound $M_u$ for the mass of the system,
\begin{equation}
\label{mstar}
M_u=N m \sqrt{1-\frac{N(N-1)^3}{8}\frac{\alpha^2}{Q^2}}.
\end{equation}
For the ground state of a boson system with $N\gg 1$ ($\alpha \ll 1$), it is easy to determine that
\begin{equation}
\label{mstar}
M_u \le \frac{D}{\sqrt{2}\, G\, m}.
\end{equation}
For $D=3$, we find $M_u \le 2.121 (G\, m)^{-1}$. This is to be compared with the result $M_u \le 1.439 (G\, m)^{-1}$ found in Ref.~\cite{rayn94}. This last bound is better, but the envelope theory can yield information about the excited states as well as fermion systems, for arbitrary values of $D$. These topics will be developed elsewhere.

Lastly, we give an application of the method to a problem in quantum mechanics with a minimal length \cite{kemp95,brau99,ques10,buis10b}. If the associated deformation parameter $\beta$ is small enough to work at first order in $\beta$, one can use the following kinetic term for a particle with a mass $m$,
\begin{equation}
\label{minx}
T(p)=\frac{p^2}{2 m} + \frac{\beta}{m}  p^4,
\end{equation}
in a nonrelativistic problem \cite{brau99}. Let us consider the $N$-body system bound by two-body harmonic potentials $V(x)=k\, x^2$. We can treat the $\beta$-terms as a perturbation. The solution of the unperturbed Hamiltonian is trivial with the system~(\ref{AFM1N})-(\ref{AFM3N}), and the use of (\ref{Epert}) gives
\begin{equation}
\label{Epbeta}
E_p=\sqrt{\frac{2\, N\, k}{m}}Q + 2\,k\,\beta\,Q^2.
\end{equation}
Note that formula (\ref{Epbeta}) is exact when $\beta=0$. For $N=2$ and $D=3$, the value of the perturbation is $2\,k\,\beta(4 n^2+l^2+4 n l+6 n+3 l+9/4)$. This compares well with the value $2\,k\,\beta(6 n^2+l^2+6 n l+9 n +4 l+15/4)$ found in Ref.~\cite{brau99} by the perturbation theory.

\section{Concluding remarks}
\label{sec:rem} 

The envelope theory is a powerful method to treat eigenvalue equations in quantum mechanics. In this paper, it is shown that it can be applied to $N$-body systems with identical particles in $D$ dimensions. One-body and two-body potentials can be considered, as well as arbitrary kinetic parts. The method is easy to implement since it reduces to find the solution of a transcendental equation. In the most favorable cases, the approximate eigenvalue is an analytical lower or upper bound. In the less favorable situations, a non-variational numerical approximation can be computed, which is often interesting for $N$-body problems which are always difficult to solve.

Several applications of the envelope theory to quantum mechanical $N$-body systems are given here. It seems that quite reliable results can be obtained. The method has also been applied to the study of spin contributions for baryons in the limit of a great number of colors \cite{buis12}. But, the applications seem potentially numerous in various domains of physics. It could be interesting to extend the method to other types of interaction, like many-body forces, or to systems with two, or more, different types of particles. 

\section*{Acknowledgment}

C.S. would thank Fabien Buisseret for useful discussions. 

\appendix

\section{Semiclassical interpretation}

Though, the envelope theory is a full quantum calculation, a semiclassical interpretation of the main equations is possible. For $N=1$ and 2, it is given in Ref.~\cite{sema12} for $D=3$, but the precise value of $D$ does not matter. 

We develop here the interpretation in the general case. Let us assume that the $N$ particles are in circular motion with the same momentum $p_0$ at a distance $d_0$ from the center of mass, each particle being at an angular distance of $2\pi/N$ from its neighbors. A semiclassical quantification of the total orbital angular momentum gives $L + DN/2=N d_0 p_0=r_0 p_0$ with $r_0=N\, d_0$. This is very similar to (\ref{AFM2N}), with the radial excitations absent from $Q$.

The total kinetic energy is $N\,T(p_0)$ and the total potential energy for the one-body interaction is $N\,U(d_0)=N\,U(r_0/N)$. The mean distance $e_0$ between two particles on the circle is 
\begin{equation}
\label{e0}
e_0=\frac{1}{N-1}\sum_{i=1}^{N-1} 2\, d_0\, \sin \frac{i\, \pi}{N} = \frac{r_0}{C_N}\cot \frac{\pi}{2 N}.
\end{equation}
Since the relative difference between $e_0$ and $r_0/\sqrt{C_N}$ is at worst 10\%, we find that the total potential energy for the two-body interaction is around $C_N\, V(r_0/\sqrt{C_N})$. We recover (\ref{AFM1N}) by adding these three contributions. It is worth mentioning a supplementary funny result. If we assume that $D \ge N-1$, the $N$ particles can be put on a hypersphere of radius $d_0$ at the vertices of a regular simplex. The distance $e$ between two particles is a constant which is the length of the edge of the simplex, with $d_0=e \sqrt{(N-1)/(2N)}$ \cite{coxe}. So, we have $e=r_0/\sqrt{C_N}$ and the two-body potential energy is exactly $C_N\,V(r_0/\sqrt{C_N})$. The quantum mechanics with the symmetrization procedure predicts a mean distance between the particles which is not possible to achieve in a (semi)classical way in our world when $N>3$. 

Since the motion is circular, each particle experiences a centripetal force $F_c$. With the definition of the effective mass proposed in Ref.~\cite{arie92}, this force is given by \cite{sema12}
\begin{equation}
\label{Fc}
F_c = \frac{p_0}{T'(p_0)} \frac{T'(p_0)^2}{d_0} = N p_0 T'(p_0) \frac{1}{r_0} .
\end{equation}
This force is driven by the potentials. The one-body contribution is centripetal and is simply given by $F_1=\left. U'(x)\right|_{x=d_0}=U'(r_0/N)$. For the two-body contribution, we must take into account that the forces act in various directions. An approximate computation of the centripetal force due to the $N-1$ other particles gives
\begin{equation}
\label{F2}
F_2 \approx \left. V'(x)\right|_{x=e_0} \sum_{i=1}^{N-1} \sin \frac{i\, \pi}{N} = V'(e_0)\cot \frac{\pi}{2 N}. 
\end{equation}
So, $F_2$ is around $\sqrt{C_N}\,V(r_0/\sqrt{C_N})$. Writing $F_c=F_1+F_2$, we recover (\ref{AFM3N}). In the situation $D \ge N-1$ described above, each particle feels the same potential $V(r_0/\sqrt{C_N})$ from all other particles and experiences the same force. But, in the computation of the centripetal contribution, we must take into account that the direction of another particle makes an angle $\alpha$ with the radius. This angle is such that $\cos \alpha = \sin \phi$ with $\sin \phi = \frac{e}{2 d_0} =\sqrt{\frac{N}{2(N-1)}}$ \cite{coxe}. In this case $F_2$ is given by 
\begin{equation}
\label{F2b}
F_2 = (N-1) \cos \alpha \left. V'(x)\right|_{x=e} = \sqrt{C_N}\,V'\left(\frac{r_0}{\sqrt{C_N}}\right),
\end{equation}
which is the exact contribution in (\ref{AFM3N}).

\end{document}